\documentclass[prb,aps]{revtex4}
\usepackage{color,graphicx,bm,amsmath}

\begin{document}

\newcommand{\note}[1]{\textcolor{red}{#1}}

\newcommand{\ir}{\ensuremath{\mathit{i}}}
\newcommand{\sr}{\ensuremath{\mathit{s}}}
\newcommand{\pr}{\ensuremath{\mathit{p}}}

\newcommand{\pa}{\ensuremath{|\pr|^2}}
\newcommand{\sa}{\ensuremath{|\sr|^2}}
\newcommand{\ia}{\ensuremath{|\ir|^2}}

\newcommand{\xfp}{\ensuremath{|\mbox{\small{\it X}}_{\pr}|}}
\newcommand{\xfs}{\ensuremath{|\mbox{\small{\it X}}_{\sr}|}}
\newcommand{\xfi}{\ensuremath{|\mbox{\small{\it X}}_{\ir}|}}
\newcommand{\cfp}{\ensuremath{|\mbox{\small{\it C}}_{\pr}|}}
\newcommand{\cfs}{\ensuremath{|\mbox{\small{\it C}}_{\sr}|}}
\newcommand{\cfi}{\ensuremath{|\mbox{\small{\it C}}_{\ir}|}}

\newcommand{\third}{\ensuremath{\frac{1}{3}}}
\newcommand{\half}{\ensuremath{\frac{1}{2}}}
\newcommand{\quarter}{\ensuremath{\frac{1}{4}}}
\newcommand{\degree}{\ensuremath{^{\circ}}}
\newcommand{\xnl}{\ensuremath{\chi^3}}
\newcommand{\is}{\ensuremath{\rm i}}

\newcommand{\eq}[1]{Eq.\ref{eq:#1}}
\newcommand{\fig}[1]{Fig.\ref{fig:#1}}

\title{Effects of Polariton Energy Renormalisation in the Microcavity 
Optical Parametric Oscillator}

\author{D. M. Whittaker}

\affiliation{Department of Physics and Astronomy, University of
Sheffield, Sheffield, S3 7RH, United Kingdom.}

\date{\today}

\begin{abstract}
The CW microcavity optical parametric oscillator (OPO) state is
investigated using a theoretical treatment which includes the
contributions of the pump, signal and idler populations to the
renormalisation of the polariton energies (the `blue-shift'). The
theory predicts the pumping conditions under which the OPO switches
on, showing that a pump angle $>\sim10\degree$ is required, but there
is no particular significance to the `magic-angle' where pump, signal
and idler are all on resonance. The signal and idler renormalisation
contributions also causes the signal emission to be driven towards the
normal direction as the pump power increases above threshold.
\end{abstract}


\maketitle

The behaviour of a resonantly pumped semiconductor microcavity can be
best described using the terminology of the optical parametric
oscillator (OPO): with pumping above threshold at finite angle, a
coherent signal appears, close to the normal direction, accompanied by
an idler on the other side of the
pump.\cite{stevenson2000,baumberg2000} However, this is an unusual
OPO, as the nonlinearity is \xnl, rather than the typical
$\chi^2$. One consequence is that there is a self-interaction, the
`blue-shift', which can be thought of as a
renormalisation of the polariton energies. This renormalisation
obviously depends on the pump, signal and idler populations. However,
although the term has been discussed in previous theoretical work, its
effect on the OPO state have been ignored, or only the pump
contribution has been considered. The purpose of the present paper is
to show that, when the renormalisation is properly treated, a number
of puzzling aspects of the experimental results can be explained.

Recent experiments\cite{butte2003} have shown that the CW-OPO can be
switched on for a range of pump angles and energies, from 10\degree\/ to
at least 24\degree, with relatively small changes in threshold. This
CW situation contrasts with the ultra-fast behaviour, where the the
response is strongly peaked about a `magic-angle', of $\sim
16\degree$, when the pump, signal and idler all lie on the polariton
dispersion.\cite{savvidis2000,erland2000} By investigating the stability of a
state in which only the pump polariton is occupied, the theoretical
treatment with the self-interaction provides a prediction of values of
the pump angle, power and energy at which the OPO can occur. In fact,
the OPO is not the only instability of the system: a simple bistable
behaviour can also be obtained, as observed experimentally by Baas
{\em et al}\/\cite{baas2003} for normal incidence pumping. The theory
predicts that for small pump angles, only bistability occurs, while
for angles greater than $\sim 10\degree$ the OPO switches
on. Furthermore, the OPO threshold is found to vary smoothly with pump
angle and energy, giving no special significance to the magic angle.

The other aspect of the experimental behaviour which is addressed here
concerns the direction of the emerging signal, which is always found
to be within a few degrees of the surface normal, whatever the angle
of the pump. This is surprising because, if the signal angle were
simply determined by the requirement that the mismatch of the signal
and idler energies from the polariton dispersion is minimised, a wider
range of angles would be obtained, depending on the pump energy and
angle. The present theory predicts that this will in fact happen, but
only very close to the threshold: at higher pump powers the
renormalisation contributions from the signal and idler switch off the
high angle OPO and pull the small angle states into resonance. In a
theory without these contributions, the pump population is pinned to
its threshold value, so nothing can change at higher pump powers.

The first theoretical models of the microcavity
OPO\cite{ciuti2000,whittaker2001} showed that the polariton energy
renormalisation arises naturally from a description in terms of
exciton-exciton scattering\cite{ciuti2000} or a \xnl\/
nonlinearity.\cite{whittaker2001} Although the renormalisation was
derived, its effects were not included in the solution for the OPO
states. The renormalisation of the pump polariton state was shown by
Baas {\it et al}\cite{baas2003} to explain the observed bistability,
but this treatment was not extended to consider the OPO state.
Gippius {\it et al}\/\cite{gippius} discuss the bistability and OPO
instability, making comparisons with a numerical model, including a
continuum of modes, from which emerges the property of emission close
to the normal. The present work is an analytic treatment which
provides a physical explanation for the numerical and experimental
results, by including the contributions to the renormalisation from
the signal and idler populations in the description of the OPO.

The treatment assumes that the pump beam is spatially uniform, and the
pump, signal and idler fields are simple plane waves. Although this is
not fully realistic, as all experiments use a finite
spot, it should give reasonable results for illumination with gradual
spatial intensity variations. As will be discussed below, some
differences between the theoretical predictions and experiment are
probably due to the effects of spatially inhomogeneous pumping.  It
should also be noted that even for uniform pumping, it is possible
that there are solutions with spatial structure to the internal
fields.

The theoretical expressions derived in this paper are fully
two-dimensional, valid for any in-plane wave-vectors consistent with
momentum conservation. However, for simplicity, the results presented
in the figures assume the pump, signal and idler lie on a line passing
through $k=0$ in wave-vector space, and the excitations considered in
the stability treatments are similarly confined. Calculations have
also been made without this restriction, and stable OPO states which
do not quite fall on such a radial line can be found. Similarly,
extending the stability treatment to consider all excitations tends to
reduce slightly the stabilty ranges of the solutions described.

The remainder of the paper is developed in the following way: In
Section \ref{sec:pump}, the state with only pump polariton mode
occupied is described, and its stability is investigated, leading to a
criteron for determining the conditions under which the OPO  switches
on.  In Section \ref{sec:opo}, the model of the OPO state with finite
amplitude is derived, showing how the emission is pulled towards the
normal direction. Finally, in Sections \ref{sec:stability} and
\ref{sec:discussion} the OPO state stability is investigated, and the
various predictions of the model are discusssed.

\section{The Pump State}
\label{sec:pump}

This section describes a state in which only a single,
pumped polariton mode is populated. The
conditions under which this state becomes unstable are derived, and
criteria are obtained for determining whether a simple bistability
results, or the OPO switches on.

\begin{figure}
\begin{center}
\mbox{
\includegraphics[scale=0.5]{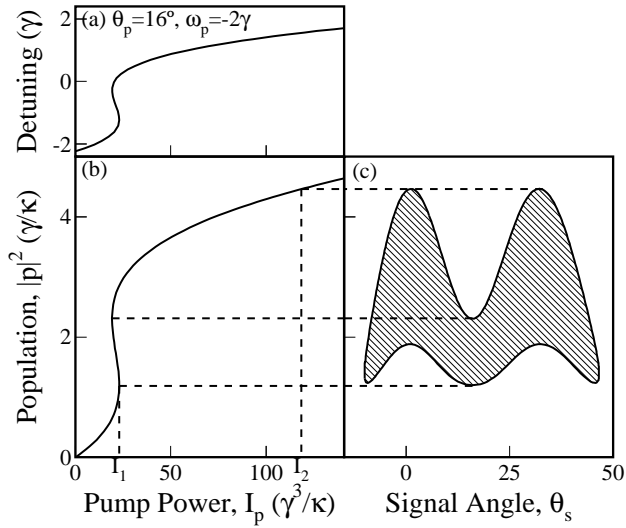}
}
\end{center}

\caption{
The pump-only state for $\theta_{\pr} = 16\degree$, $\omega_{\pr}=-2
\gamma$.  (a) Detuning of the pump relative to the polariton energy,
including the renormalisation contribution.  (b) S-curve showing the
relationship between the pump polariton population \pa\/ and the
external pump power $I_p$. (c) Values of
\pa\/ (shaded region) for which the pump-only state is unstable, as a
function of the signal angle, $\theta_s$.  $I_1$ and $I_2$ are the pump
powers corresponding to the lowest and highest unstable values of
\pa\/ for any $\theta_s$: the OPO switches on when $I_2>I_1$.
 }
\label{fig:scurve}
\end{figure}

The pump mode is assumed to be a plane wave 
$\phi_{\pr}=(\pr/\xfp^2) \exp{\is (k_{\pr} x-\omega_{\pr} t)}$, 
where the amplitude, \pr, satisfies:\cite{whittaker2001}
\begin{equation}
\label{eq:bistable}
 \frac{1}{\xfp^2}(\omega_0(k_{\pr})-\is \gamma_{\pr} - \omega_{\pr})
\pr +\kappa {\pa} \pr=\frac{\cfp}{\xfp} f_{\pr} .
\end{equation}
Here, $\omega_0(k_{\pr})$ is the lower branch dispersion at the pump
wave-vector, $\gamma_{\pr}$ its width, \cfp\/ and \xfp\/ the cavity
and exciton amplitudes (Hopfield factors), and $f_p$ the external pump
amplitude. The non-linear term, $\kappa \pa\pr$ represents the self
interaction of the pump polaritons. By appropriate scaling of the
fields,\cite{whittaker2001} the coefficient $\kappa$ can be made equal
to unity, which forthwith is taken to be the case. The use of this
single-branch form assumes that the polariton coupling is
sufficiently large that the two branches are not mixed
by the non-linear terms.

The solution of the cubic \eq{bistable} has been discussed in
Ref(\onlinecite{baas2003}): when the detuning
$\omega_{\pr}-\omega_0(k_{\pr}) > \sqrt{3} \gamma_{\pr}$, the cubic
has two turning points, resulting in a characteristic S-curve, as
shown in
\fig{scurve}(b)\cite{parameters}. For pump-powers falling between the two turning
points, the system is bistable (the branch with negative slope is
unstable). \fig{scurve}(a) shows how this bistability is intimately
linked with the zero-crossing of the pump detuning, as it changes due
to the blue-shift contributions. For negative detunings, the shift
pulls the polariton feature towards resonance with the pump, causing
the population to grow super-linearly. However, when the detuning
becomes positive, the polariton is pushed away from resonance.  The
pump polariton population must lie on this curve if only the pump mode
is significantly occupied. However, if other modes have finite
occupation, as in the OPO state, this restriction no longer
applies. Hence the conditions under which the OPO switches on can be
found by determining which parts of the pump-only curve are
unstable.

The stability of the pump-only state is determined by considering the
spectrum of small excitations with frequency $\omega$ and wave-vector
$q$, of the form
\begin{equation}
 \phi_p=e^{\is(k_{\pr} x- \omega_{\pr} t)} \left[ p + \sr e^{-\is(q x-\omega t)}
 + \ir e^{\is(q x-\omega^* t)} \right]
\label{eq:pump}
\end{equation}
which describes signal and idler modes with amplitudes \sr, \ir, and
wave-vector/energies ($k_{\sr}=k_p-q$, $\omega_{\sr}=\omega_{\pr}-\omega$) and
($k_{\ir}=k_p+q$, $\omega_{\ir}=\omega_{\pr}+\omega$). Expanding to first
order in \sr\/ and \ir, this gives
\begin{subequations}
\begin{eqnarray}
\label{eq:signal}
 \frac{1}{\xfs^2}(\omega_0(k_s)-\is \gamma_{\sr} - \omega_{\pr}+\omega) \sr
 +2 \pa \sr + \pr^2 \ir^* \!\! &=& \!\! 0\\
\label{eq:idler}
 \frac{1}{\xfi^2}(\omega_0(k_i)-\is \gamma_{\ir} - \omega_{\pr}-\omega^*) \ir
 +2 \pa \ir + \pr^2 \sr^* \!\! &=& \!\! 0
\end{eqnarray}
\end{subequations}
Considering these equations as an eigenproblem for the
amplitudes \sr\/ and $\ir^*$, the complex eigenvalues, $\omega$, are given by
the condition that the determinant of the coefficients is zero, that is
\begin{eqnarray}
\label{eq:simpledispersion}
&(& \!\! \omega_0(k_s)-\is \gamma_{\sr} - \omega_{\pr}+\omega+2\xfs^2 \pa)
\times
\\ \nonumber
&(& \!\! \omega_0(k_{\ir})+\is \gamma_{\ir} - \omega_{\pr}-\omega+2\xfi^2 \pa)
=\xfs^2 \xfi^2 |p|^4
\end{eqnarray}
The threshold for instability corresponds to Im$\{\omega\}=0$, which
occurs when
\begin{equation}
\label{eq:threshold}
\Delta^2 + \Gamma^2
= \frac{\Gamma^2}{\gamma_{\sr} \gamma_{\ir}} \xfs^2 \xfi^2 |p|^4, 
\end{equation}
where $\Gamma=\gamma_{\sr}+\gamma_{\ir}$ and 
\begin{equation}
\label{eq:mismatch}
\Delta=\omega_0(k_s) + \omega_0(k_{\ir}) - 2 \omega_{\pr}
+2(\xfs^2+\xfi^2) \pa
\end{equation}
is the mismatch from the resonance condition $2
\omega_{\pr}=\omega_0(k_s) + \omega_0(k_{\ir})$, modified by the
blue-shift. Eqs. (\ref{eq:threshold}) and (\ref{eq:mismatch}) provide a
quadratic condition for \pa\/ at the boundary of the instability region.

\fig{scurve}(c) shows the instability region, plotted as a function of the
signal angle $\theta_{\sr}$. For $\theta_{\sr}=\theta_{\pr}$, the
extrema of the instability region correspond to the two turning points
of the S-curve, so this analysis correctly predicts that the
negative-slope portion of the curve is unstable. However, for the
pumping conditions shown, the total extent of the instability region
for all $\theta_{\sr}$ is significantly greater: converting the
extremal \pa\/ values to pump powers gives a range between $I_1$ and
$I_2$ on \fig{scurve}(b), for which the pump-only state is unstable.

The discussion above provides a criterion for determining whether
the OPO state turns on for a given pump angle and energy: if there is
a range of pump powers $I_p$ for which the pump-only state is
unstable, something more complicated, presumably the OPO, must turn
on in between. In terms of \fig{scurve}(b), this means that $I_2>I_1$. If, on the
other hand, $I_2<I_1$, the system can jump straight
from the lower branch of the S-curve to a stable state on the upper
branch, and a simple bistability occurs. Of course, this does not rule
out the possibility of the OPO switching on, but numerical experiments
similar to those described in Ref.(\onlinecite{gippius}) only
produce bistability.\cite{numerics} A further possibility is that there is no
unstable region, and \pa\/ simply rises smoothly with increasing
pump-power. Note that the OPO state does not require a bistable
S-curve, as the unstable region of \fig{scurve}(c) does not have to
include $\theta_{\sr}=\theta_{\pr}$. 

\begin{figure}
\begin{center}
\mbox{
\includegraphics[scale=0.45]{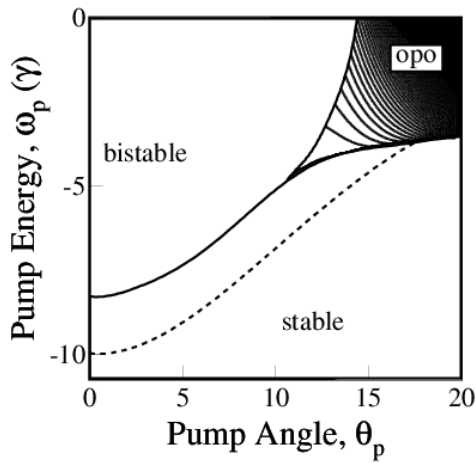}
}
\end{center}
\caption{
Divison of the ($\theta_p$,$\omega_{\pr}$) plane into sectors of different
behaviour. In the `stable' region, as the pump power, $I_p$\/ rises,
the pump population, \pa, increases smoothly. In the bistable region,
there is a discontinuous jump in \pa, but no other state becomes
occupied. The contours in the OPO region show the variation in the
threshold, increasing in steps of $10 (\gamma^3/\kappa)$ from the bottom left to the
top right. The bare polariton dispersion is shown as a dashed line.
}
\label{fig:threshold}
\end{figure}

In \fig{threshold}, the sectors of ($\theta_{\pr}$, $\omega_{\pr}$)
corresponding to these three types of behaviour are delineated. The
figure shows that there is a minimum pump angle, $\sim
10\degree$, below which it is not possible to turn the OPO on. In the
OPO sector, the threshold rises monotonically with increasing
$\theta_{\pr}$ and $\omega_{\pr}$, and there is no special
significance to the `magic angle', $\theta_{\pr} \sim 16\degree$, at
which the pump, signal and idler can all be made resonant with the
lower branch dispersion. These conclusions are broadly consistent with
the experimental results of Butt\'e {\it et al}\/\cite{butte2003}.

\begin{figure}
\begin{center}
\mbox{
\includegraphics[scale=0.5]{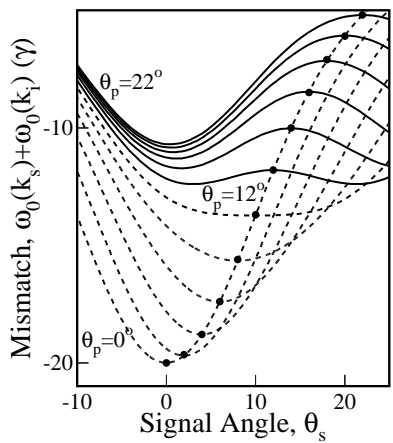}
}
\end{center}

\caption{
The bare mismatch, $\omega_0(k_{\sr})+\omega_0(k_{\ir})$, plotted as a
function of signal angle, $\theta_{\sr}$, for different values of pump
angle, $\theta_{\pr}$. The filled circles indicate the points where
which $\theta_{\sr}=\theta_{\pr}$. For
$\theta_{\pr}<12\degree$, the dashed curves, these points are
minima and there is no OPO. For $\theta_{\pr}\ge 12\degree$, they are
maxima, and the OPO switches on. In this regime, the minimum mismatch
always occurs close to $\theta_{\sr}=0\degree$.
 }
\label{fig:miss}
\end{figure}

The key property which determines whether the OPO switches on appears
to be the presence of a minimum in the higher threshold for
$\theta_{\sr}=\theta_{\pr}$, as in \fig{scurve}(c). If this is the
case, the upper threshold, $I_2$, is shifted above the higher knee in
the S-curve, providing a gap in the stability of the pump-only curve.
In \fig{miss}, it is shown that this correlates with the presence of a
maximum in the $\theta_{\sr}$ dependence of the bare mismatch when
$\theta_{\sr}=\theta_{\pr}$: for $\theta_{\pr}<10\degree$, instead
there is a minimum for this value of $\theta_{\sr}$.

\section{The OPO States}
\label{sec:opo}

The previous section described the OPO states at threshold, when the
signal and idler populations are still zero.
In this section, the treatment is extended to include
contributions of finite signal and idler populations to the dispersion
renormalisation.  This model is solved to obtain the variation of the
pump, signal and idler populations above threshold for a range of
signal angles. 

Eqs. (\ref{eq:pump}), (\ref{eq:signal}) and (\ref{eq:idler}) are only
accurate to first order in \sr\/ and \ir. For finite signal and idler
populations, they are modified to include the renormalisation effects
of the signal and idler populations, and the depletion of the pump:
\begin{widetext}
\begin{subequations}
\begin{eqnarray}
\label{eq:fullpump}
 \frac{1}{\xfp^2}(\omega_0(k_p)-\is \gamma_{\pr} - \omega_{\pr}) \pr 
 +(\pa + 2 \sa + 2 \ia) \pr  + 2 \pr^*s\ir
  \!\! &=&  \!\! \frac{\cfp}{\xfp} f_{\pr} \\
\label{eq:fullsignal}
 \frac{1}{\xfs^2}(\omega_0(k_s)-\is \gamma_{\sr} - \omega_{\sr}) \sr
 +(2 \pa +\sa + 2\ia) \sr + \pr^2 \ir^* \!\! &=& \!\! 0\\
\label{eq:fullidler}
 \frac{1}{\xfi^2}(\omega_0(k_{\ir})-\is \gamma_{\ir} - \omega_{\ir}) \ir
 +(2 \pa +2 \sa + \ia) \ir + \pr^2 \sr^* \!\! &=& \!\! 0.
\end{eqnarray}
\end{subequations}
\end{widetext}
Note the different form for the renormalisation of each state: the
self-interaction term has only half the strength of each cross
term. In the OPO state the renormalisation does {\em not} produce a
uniform blue-shift of the polariton dispersion.

With the condition that $\omega_{\sr}$ and $\omega_{\ir}=2
\omega_{\pr}-\omega_{\sr}$ must be real,
Eqs.(\ref{eq:fullsignal},\ref{eq:fullidler}) require
\begin{equation}
\omega_{\sr}= \frac{\gamma_i}{\Gamma} \left[ \omega_0(k_s)+\xfs^2 ( 2 \pa +\sa + 2
\ia) \right]
\end{equation}
and for this value of $\omega_{\sr}$
\begin{equation}
\label{eq:ratio}
 \frac{\sa}{\ia} = \frac{\xfs^2 \gamma_i}{\xfi^2 \gamma_{\sr}}
\end{equation}
Then \pa, \sa\/ and \ia\/ satisfy \eq{threshold} with the mismatch
modified to include the signal and idler contributions to the
renormalisation,
\begin{eqnarray}
\label{eq:fulldelta}
\Delta &=& \omega_0(k_s) + \omega_0(k_i) - 2 \omega_{\pr}
 + \xfs^2 (2 \pa +\sa +2 \ia) \nonumber \\  & & \mbox{} +  \xfi^2(2 \pa + 2 \sa +\ia),
\end{eqnarray}
which, using \eq{ratio}, represents a biquadratic
equation connecting \pa\/ and \sa. A further relationship between
the two variables can be obtained by solving \eq{fullidler}
for \ir\/ in terms of $\sr^*$, and substituting in the pump equation
(\ref{eq:fullpump}) to get
\begin{eqnarray}
 \frac{1}{\xfp^2}(\omega_0(k_p)-i \gamma_{\pr} - \omega_{\pr}) p 
 +(\pa + 2 \sa + 2 \ia) p \nonumber \\ 
 -2 \xfi^2 \frac{\Gamma}{\gamma_i}
 \frac{\sa \pa}{\Delta + \is \Gamma}
  = \frac{\cfp}{\xfp} f_p .
\end{eqnarray}
These two relationships can readily be solved to obtain the values of
\pa\/ and \sa\/ in the OPO states, including the contributions of the
signal and idler populations to the blueshifts.

\begin{figure}
\begin{center}
\mbox{
\includegraphics[scale=0.55]{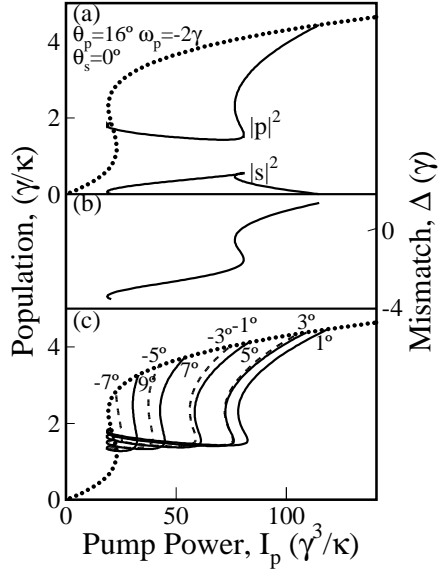}
}
\end{center}

\caption{
The OPO states including signal and idler renormalisation
contributions.
(a) Relationship between the pump population, \pa, and pump power,
$I_{\pr}$, for signal angle $\theta_{\sr}=0$, and the
corresponding signal population \sa.
(b) The mismatch $\Delta$ from \eq{fulldelta} as a function of
$I_{\pr}$ for $\theta_{\sr}=0$.
(c)  The pump population variation, as in (a), for a number of values
of $\theta_{\sr}$. The curves for negative values of $\theta_{\sr}$ are shown
dashed for clarity.
}
\label{fig:opo}
\end{figure}

Such solutions for the OPO state are shown in \fig{opo}. Above
threshold, there is generally a region where the pump population, \pa,
remains nearly flat, while the signal, \sa, increases with pump
power. In a treatment with just the pump contribution to the
renormalisation, \pa\/ would be completeley flat for all powers above
threshold. However, in the present model, a power is reached where
the flat behaviour ceases, and
\pa\/ increases rapidly towards the pump-only line, at which point
the OPO switches off, with \sa\/ falling to zero. Examining the curves
for different pump angles $\theta_{\sr}$ reveals that the maximum
$I_{\pr}$ at the turn-over occurs when $\theta_{\sr} \sim 1\degree$,
and it decreases rapidly on either side, so at $-7\degree$ and
$+9\degree$ there is only a small range of $I_{\pr}$ in which the OPO
state exists.

The origin of this behaviour can be understood by considering the
mismatch $\Delta$, from \eq{fulldelta}, which is zero when the pump
and the renormalised signal and idler satisfy the triple resonance
condition.  As is shown in \fig{opo}(b), $\Delta$ is negative in the
flat region, where the increasing blue-shift is thus pulling the OPO
towards resonance, and helping the signal to increase. The turn-over
occurs around the pump power where $\Delta$ becomes positive, so
increases in the blue-shift drive the system away from resonance, and
the OPO rapidly turns off. The dependence on signal angle is therefore
a consequence of the different values of the bare mismatch
$\omega_0(k_{\sr})+\omega_0(k_{\ir})-2
\omega_{\pr}$, which, because of the dispersion shape, always has its
minimum close to $\theta_{\sr}=0$. Indeed, it can be seen from
\fig{miss} that this occurs at approximately
$+1\degree$ for $\theta_{\pr}=16\degree$, and changes very little over
the range of pump angles for which the OPO switches on.
Thus, though the signal
angle with the smallest bare-mismatch has a lower threshold, when the
pump power is increased the
growing blue-shift quickly switches that state off and the
signal moves towards $\theta_{\sr}=0$.

\section{OPO stability}
\label{sec:stability}

The solutions for the OPO states described in the previous section are
only of physical relevance if they are stable. This section describes a
linear stability analysis of the OPO, and shows that there are stable
solutions, but they become unstable when the pump population curves on
\fig{opo} turn over.

The stability analysis for the OPO proceeds in a similar manner to
that for the pump-only state. The spectrum of small excitations about
the OPO state is calculated using a linearised expansion of
fluctuations in the pump, signal and idler modes. For an excitation
wave-vector $q$ and frequency $\omega$, these fluctuations satisfy a
set of coupled equations, from which the dispersion is obtained using
the condition that the determinant of the coefficients is zero. Thus
\begin{widetext} 
\begin{equation}
\label{eq:dispersion}
\left|
\begin{array}{cccccc}
d_{\pr} & \pr^2+2\sr\ir & 2(\pr\sr^*+\pr^*\ir) & 2\pr\sr
 & 2(\pr^*\sr+\pr\ir^*) &  2\pr\ir \\
\pr^{*2}+2\sr^*\ir^* & \overline{d}_{\pr} & 2\pr^*\sr^* &
2(\pr^*\sr+\pr\ir^*) & 2\pr^*\ir^* & 2(\pr\sr^*+\pr^*\ir) \\
2(\pr^*\sr+\pr\ir^*) & 2\pr\sr & d_{\sr} & \sr^2 & 2\sr\ir^* &
\pr^2+2\sr\ir \\
2\pr^*\sr^* & 2(\pr\sr^*+\pr^*\ir) & \sr^{*2} & \overline{d}_{\sr} &
\pr^{*2}+2\sr^*\ir^* & 2\sr^*\ir^* \\
2(\pr\sr^*+\pr^*\ir) & 2\pr\ir & 2\sr^*\ir & \pr^2+2\sr\ir & d_{\ir} &
\ir^2 \\
2\pr^*\ir^* & 2(\pr^*\sr+\pr\ir^*) & \pr^{*2}+2\sr^*\ir^* & 2\sr\ir^*
& \ir^{*2} & \overline{d}_{\ir} 
\end{array}
\right|=0
\end{equation}
where
\begin{eqnarray}
d_{\pr} &=& \frac{1}{\xfp^2}(\omega_0(k_{\pr}-q) - \is \gamma_{\pr} -
\omega_p + \omega) + 2(\pa + \sa +\ia) - \pa \nonumber \\
\overline{d}_{\pr} &=& \frac{1}{\xfp^2}(\omega_0(k_{\pr}+q) + \is \gamma_{\pr} -
\omega_p - \omega) + 2(\pa + \sa +\ia) - \pa 
\end{eqnarray}
with equivalent expressions for $d_{\sr}$, $\overline{d}_{\sr}$
etc. The terms in the determinant are, in order of the columns, 
the coefficients multiplying the amplitudes of the fluctuations in
\pr, $\pr^*$, \sr, $\sr^*$, \ir\/ and $\ir^*$.

\end{widetext}

\begin{figure}
\begin{center}
\mbox{
\includegraphics[scale=0.6]{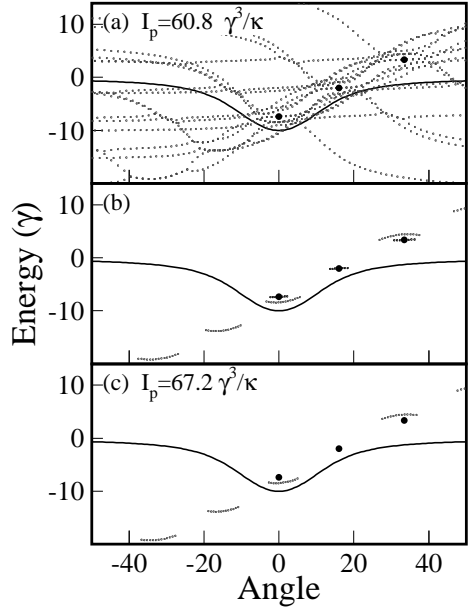}
}
\end{center}
\caption{
Real part of dispersion curves for excitations from the OPO state with
$\theta_p=16\degree$, $\omega_{\pr}=-2 \gamma$ and
$\theta_s=0\degree$. (a) shows all the branches for the unstable state
with $I_p=60.8 \gamma^3/\kappa$, (b) the branches which
are actually unstable. (c) shows the unstable branches for the
nominally `stable' state with $I_p=67.2 
\gamma^3/\kappa$, as discussed in the text. The solid line is the bare
polariton dispersion and the large points are the ($\theta$,$\omega$)
locations of the pump, signal and idler.  }
\label{fig:stability}
\end{figure}

In \fig{stability} the eigenvalues calculated from
\eq{dispersion} are plotted. \fig{stability}(a) shows the real parts of the
roots for all the calculated eigenmodes. The calculated energies,
$\omega$, are the excitations from the OPO state, which has three
energies and wave-vectors ($\omega_{\pr}$,$k_{\pr}$),
($\omega_{\ir}$,$k_{\ir}$) and ($\omega_{\ir}$,$k_{\ir}$). Thus, in a
similar manner to the discussion concerning \eq{pump}, each excitation
corresponds to six absolute energies and wave-vectors, ($k_{\pr} \pm
q$, $\omega_{\pr} \pm \omega$), ($k_{\sr} \pm q$, $\omega_{\sr} \pm
\omega$) and ($k_{\ir} \pm q$, $\omega_{\ir} \pm \omega$).
These are the values at which the excited states can be created from,
or transformed into an external photon, and it is these energies and
wave-vectors which are plotted on the figure. The resulting dispersion
curves, which are highly complicated, could be observed as parametric
luminescence in the presence of the OPO state; indeed, analogous
features have been observed in ultrafast pump-probe
measurements.\cite{savvidis2001}

\fig{stability}(b) shows only the modes which are unstable, that is
Im($\omega$)$>0$. For this pump power there are two types of unstable
excitation: small wave-vector modes, represented by the short, near
horizontal, sections of dispersion passing through the pump, signal and
idler, and much larger wave-vector modes which give rise to the longer
arcs. For example, the long arc close to the idler is actually an
excitation from the signal, with wave-vector $q \sim k_{\ir} -
k_{\sr}$, and it is paired with the arc at $\sim \,-30\degree$.

\fig{stability}(c) corresponds to a slightly higher pump power than
(b), where the small wave-vector instability has disappeared. However
the large wave-vector modes remain, and indeed are always found in the
present calculations, suggesting that the simple OPO state is never
really stable. In fact, this is known to be true: in the experiments
of Tartakovskii {\it et al},\cite{tartakovskii2002} the OPO state is
found to be accompanied by additional weak satellites, $s'$ and $i'$,
at $k_{s'}=2k_{\sr}-k_{\pr}$ and $k_{i'}=2k_{\ir}-k_{\pr}$,
corresponding to the scattering processes $p+p
\rightarrow s'+i'$, $s+i \rightarrow s'+i'$, $p+s \rightarrow  s'+i$
and $p+i \rightarrow s+i'$. Such satellites can also be seen in the
numerical simulations of Ref.(\onlinecite{gippius}). These particular
states are selected because the corresponding $q$ is equal to
$k_{\pr}-k_{\sr}$, a resonance condition which is not accounted for in the
derivation of \eq{dispersion}. However, the  comparison suggest that the
instability associated with the large wave-vector modes is benign, and
does not grow very large before being limited by non-linear terms.
Distinguishing such instabilities from the more catastrophic variety
is not possible within a linear stability analysis of the type
performed here, which only shows that an infinitesimally small
fluctuation in one of these modes will initially grow. 

\begin{figure}
\begin{center}
\mbox{
\includegraphics[scale=0.7]{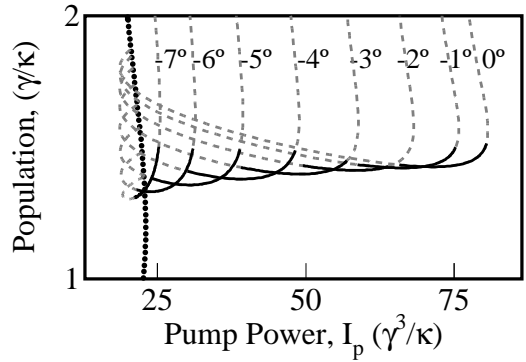}
}
\end{center}

\caption{
Stable regions of \pa versus $I_p$, for various signal angles
$\theta_s$, according to the criteria discussed in the text. The solid
lines represent the stable states.
}
\label{fig:opostability}
\end{figure}

Adopting the criterion that only the small wave-vector instabilities are
important, \fig{opostability} shows the portions of the OPO curves
with negative $\theta_{\sr}$, in
\fig{opo}(c), which are stable. It can be seen that for a particular
pump power, the OPO state with the lowest pump population \pa\/ is
stable. This is consistent with the discussion in
Ref.(\onlinecite{whittaker2001}), where it was argued that if a state
with lower \pa\/ were possible, the higher \pa\/ state would be
unstable. However, in the present treatment, the stable region for a
particular angle extends beyond where it has lowest \pa. This is
because stability is concerned with the response to infinitesimal
fluctuations, and for a given pump power and signal angle, a small
fluctuation in another mode may decay, where a large excursion into
that mode would be stable.

\section{Discussion}
\label{sec:discussion}

The results presented in this paper show that a proper treatment
of the polariton renormalisation leads to a model which predicts the
pump conditions under which the microcavity OPO can be observed, and
explains the tendency for the signal emission to be close to the
surface normal at high pump powers.

The OPO threshold behaviour has been shown to be a consequence of how
the mismatch in energy between the pump, signal and idler modes,
$\Delta$ in \eq{mismatch}, changes when their energies are
renormalised by the self-interaction terms. This suggests an
explanation for the observation, in ultrafast pump-probe measurements,
that the `magic-angle' for pumping is important: when an external
probe is used the energies and angles of all three modes are
determined by the experimental geometry, so the system cannot
re-arrange itself in the same way. Hence, a strong response is only
obtained when the bare polariton dispersion is resonant with the three
modes, which gives the magic-angle condition. Indeed, as CW
experiments are always carried out with the pump positively detuned
from the bare dispersion, it is probably not surprising that the triple
resonance condition is not important.

The model also predicts that in the OPO region, there is a maximum
pump power above which the OPO switches off, as well as a lower
threshold. This behaviour is observed in numerical simulations such as
Ref.(\onlinecite{gippius}), but not in experiments.  This is probably
a consequence of the inevitably inhomogeneous pump intensity
associated with a finite excitation spot. A two dimensional numerical
model\cite{numerics} with a Gaussian excitation profile shows that, as
the power is turned up, the signal switches off at the centre of the
spot, but there is always a region further out where the intensity is
in the right range for the OPO to be active.

A further prediction is that for given pumping conditions, there are a
number of stable OPO states corresponding to different signal
angles. This suggests that the actual state obtained will depend on
exactly how the OPO is switched on, and possibly on random factors
such as noise in the system. It may also be possible to steer the
signal, by using a second probe beam to favour a particular emission
angle.

\begin{acknowledgments}
I wish to thank P.~R.~Eastham, D.~Sanvitto and M.~S.~Skolnick for helpful
contributions in discussions about this work.
I also acknowledge the financial support of the EPSRC (GR/A11601)
\end{acknowledgments}


\begin{thebibliography}{}
\bibitem{parameters} In this, and all other numerical results, 
the the relative detuning of the exciton and cavity is zero at $k=0$.
The coupling strength is chosen to be 20 times the line-width, a
typical value for the good GaAs microcavities
studied experimentally.
\end{thebibliography}
\end{document}